# Optimizing Web-Based AI Query Retrieval with GPT Integration in LangChain: A CoT-Enhanced Prompt Engineering Approach

**Wenqi Guan[1*], Yang Fang[2]**

[1*]Department of Electronic Engineering, Nanjing University, Nanjing 210023, China

[2]Department of Cybersecurity, Beijing University of Posts and Telecommunications, Beijing 102206, China

[1*]221180006@smail.nju.edu.cn, [2]1966985701@qq.com

Abstract-Large Language Models have brought a radical change in the process of remote learning students, among other aspects of educative activities. Current retrieval of remote learning resources lacks depth in contextual meaning that provides comprehensive information on complex student queries. This work proposes a novel approach to enhancing remote learning retrieval by integrating GPT-based models within the LangChain framework. We achieve this system in a more intuitive and productive manner using CoT reasoning and prompt engineering. The framework we propose puts much emphasis on increasing the precision and relevance of the retrieval results to return comprehensive and contextually enriched explanations and resources that best suit each student's needs. We also assess the effectiveness of our approach against paradigmatic LLMs and report improvements in user satisfaction and learning outcomes.

## 1. Introduction

Large Language Models have revamped the field of NLP, enabling systems to perform a wide range of language-oriented tasks, such as text generation, translation, and question-answering, among others, with great accuracy. In this respect, GPT-4o is a significant example; it possesses advanced capabilities in understanding and generating contextually appropriate responses pertaining to many fields[1]. Nonetheless, although these models demonstrate significant proficiency across various domains, they frequently face difficulties in particular applications such as remote learning retrieval, wherein accurate and contextually pertinent information is essential.

Effective remote learning is pegged on how well information is retrieved and presented to the students, matching their needs. While current retrieval systems seem competent in retrieving straightforward queries, they always break down when it comes to more complex and context-dependent questions[2]. This limitation calls for enhanced retrieval frameworks that can better capture the nuances of educational content and student inquiries.

Ghodratnama and Zakershahrak propose in their paper "Adapting LLMs for Efficient, Personalized Information Retrieval"[3] some new ways to overcome the challenges mentioned above. Dmytro and Waseem followed up with a broad array of approaches, which included the integration of Azure

Cognitive Search Retriever with GPT-4; the Canopy framework from Pinecone; different configurations of LangChain along with Pinecone and multiple language models such as OpenAI and Cohere; LlamaIndex with a hybrid search on Weaviate Vector Store; and trying out Google's RAG on Cloud Vertex AI-Search and Amazon SageMaker's RAG, with a completely new approach that combined a graph search algorithm with a language model and retrieval awareness[4]. Although these approaches yielded many valuable insights, we felt that their effectiveness and accuracy in answering academic questions could be further improved. Identifying this knowledge gap compelled us to present a new approach based on the use of GPT-4 with the LangChain framework and Chain-of-Thought (CoT) reasoning, improved by state-of-the-art prompting strategies.

LangChain is specifically designed for building LLM-driven applications and hence perfectly serves the needs of adaptive environments while processing such multi-step tasks that might be entangled in the specific complexity of text-based AI tasks. Since this design decision relies on RouterChain, it ensures that with the correct intent classification of each problem, the retrieving process would self-correctively be performed. Besides, the proposed framework also tries to make the retrieval more effective using available tools like Google Search and available APIs like Spark 4.0 Ultra. This will further help in bringing up a greater diversity of sources from which the system can draw and improve depth and precision in information retrieval. The interaction between the LLM and these extra tools results in a much stronger process of decision-making in which AI is supposed to weigh different pieces of information against each other before delivering a final well-rounded answer. Our approach will focus on the limitations of traditional retrieval systems in order to provide an answer that is far more flexible and complete. Retrieval is done in an enhanced manner on the leverage of functionalities embedded in GPT-4o and LangChain for such high, complex, and dynamic needs, which are very common within remote learning environments. The contributions within this study can be enumerated as follows:

a) We introduce a novel framework that integrates CoT reasoning to improve the accuracy and relevance of remote learning retrieval.
b) We demonstrate the effectiveness of our approach through a detailed evaluation, comparing its performance against existing methods.
c) We make the source code and datasets available on GitHub for further research (https://github.com/Guanwenqi/Langchain-Retrieval-Framework).

## 2. Related Work

Large Language Models (LLMs), such as GPT-3.5 and GPT-4, have rapidly evolved to become essential tools in various text analysis and automation tasks due to their exceptional language comprehension capabilities. The desirability of their capability to conduct a range of activities-from the classification of written content, based on predetermined parameters, to the issuance of policy recommendations in government settings is very high. However, large language models have fallen remarkably short of remembering events, incorporating the latest information, and managing domain-specific content with many tiresome problems such as hallucinations and inaccuracies[5].

Such limitations have driven some researchers towards the development of Retrieval-Augmented Generation systems, which combine large language models with external retrieval databases for improved accuracy in the relevance of responses to the context. In such a combination, the models GPT-3.5 and GPT-4 are capable of providing full-fledged responses with minimum chances of generating incorrect information[6]. This has especially been helpful in many applications that require contextualized responses with high levels of precision, like education.

Beyond retrieval augmentation, the integration of KGs with LLMs has also become a strategy through which the domain-specific limitations of the latter are addressed, with responses being deeply set in structured factual contexts[7]，which is rather helpful within highly specialized domains such as healthcare and education. Thus, creating KGs using various educational data sources, including unstructured text, databases, and APIs, has proved to be effective for improving the contextual understanding of LLMs and, consequently, these models are better at returning precise, focused answers to open-ended student questions[8].

In the context of educational applications, recent studies have explored the use of LLMs to automate content comprehension and answer student queries across varying levels of complexity. Examples range from the use of the LLM in answering multiple-choice questions in Python courses to mathematical deduction. Though the very recent GPT-4 shows great improvements in both accuracy and reasoning compared to its predecessors, fulfilling its promise to change the digital learning landscape and provide personalized and accurate support to learners. These traditional LLMs still face difficulties in resolving complex, contextually heavy questions. Therefore, recent works have shifted the focus to the importance of both PE and CoT reasoning[9]. These are crucial tools that help guide problem-solving in LLMs towards structured, coherent, and logically consistent responses.

Prompt engineering is the intentional crafting of inputs, leveraging the rich knowledge lying within models like GPT-4. Such techniques as zero-shot and few-shot prompting enable these models to carry out new tasks for which little or no previous training data existed. The zero-shot prompting relies on a well-designed prompt to instruct model behaviour without labeled examples, while a few-shot prompting takes a small number of task-specific illustrations to generate the expected input-output relationship.

Building on the zero-shot and few-shot methods, the Chain-of-Thought prompting has significantly improved the capability of LLMs for complex reasoning questions[10]. By merely soliciting models to break down problems into step-by-step logical components, CoT can construct responses that would not be merely more accurate but also a lot more comprehensible, since it helps models explain why every step is done in such a manner, further enriching the learning within the responses.

Recent breakthroughs have perfected this process through automation, such as Automatic Chain-of-Thought prompting, without the tedious process of constructing the prompting chain manually. Auto-CoT takes the approach "Let's think step by step" towards generating different chains of reasoning. In this respect, the said result brought improvements in arithmetic and symbolic reasoning performance, showing its applicability in fields requiring step-by-step evaluation[11]. Besides, CoT combined with self-consistency decoding enhanced the capacity of LLMs to solve complex queries with high accuracy by generating multiple chains of reasoning and identifying the most consistent solution among those chains. This process lets the model arrive at a more reliable answer simply by marginalizing the chain, since such a task may naturally have several equally viable solutions.

Building on these, our proposed framework embeds the GPT-based model into the LangChain framework, optimized with CoT reasoning and strategies of prompt engineering. Emphasizing retrieval precision and relevance enrichment of context, our approach aims to overcome the current limitations in remote learning systems, providing students with comprehensive, contextually enriched educational resources. This novel integration not only improves the accuracy of retrieved content but also aligns with the increasing demand for privacy-preserving AI solutions in educational settings.

## 3. Framework of the Retrieval Tool

The main objective of this work is the development of an understandable system operating with heightened efficiency by utilizing Chain-of-Thought reasoning and prompt methodologies that enhance the relevance and the precision of the response given[12, 13]. This system has been developed mainly using the LangChain framework. By developing various types of agents and making use of tools, it allows us to meet most of our needs. The proposed framework would enhance the relevance and accuracy of the results during internet searches for updated information. In such a way, a response would be complete and contextually relevant to the particular needs of end-users, such as the retrieval of an arXiv paper or summarization of textual content from web pages. A visualized structure revolving around these functionalities is shown in **Figure 1**.

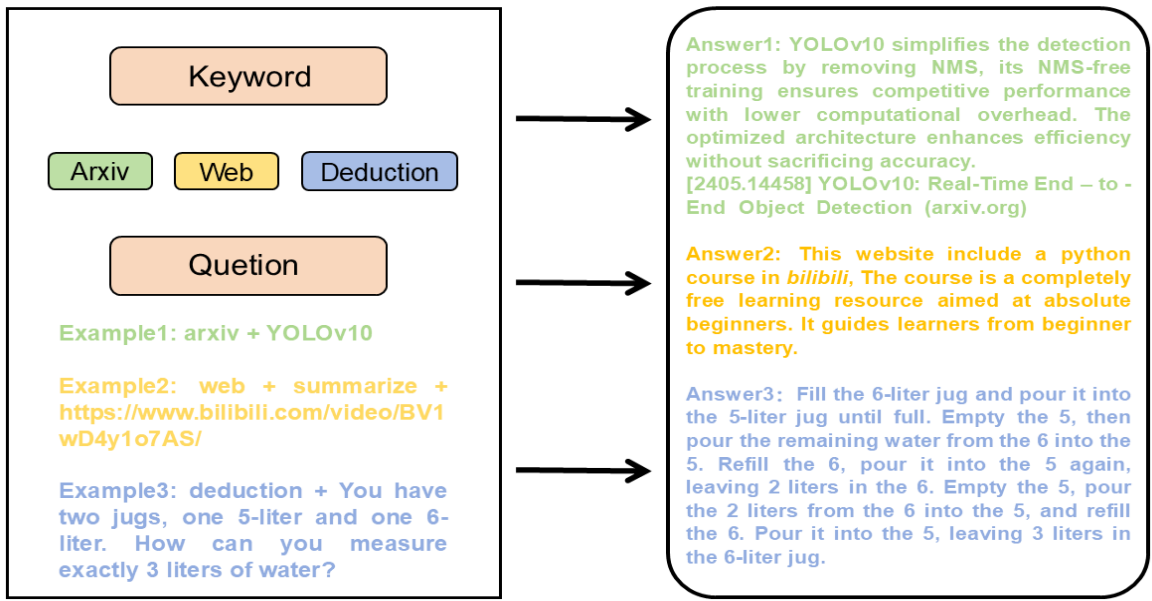

Fig.1 Structure of the specimen

*3.1. langchain*
LangChain is a very powerful framework designed for building applications that leverage large language models. Its goal is to make it easier for developers to integrate with various data sources and interact with external applications. LangChain achieves this by offering modular components and customizable chains, allowing for the creation of case-specific pipelines[14]. We use several components in our framework such as prompts, chains, agents, and memory.

*3.2. querying arXiv papers*
Proper tools and techniques in academic research can vastly help in precision and reducing the time taken for the whole process. For instance, the use of the keywords "arXiv + mamba" on a search engine would yield a set of scientific journal articles that relate to the topic under consideration. This, in fact, would not only bring in a collection of related works but also quite extended reviews of papers found.

This efficient tracing of literature is quite important in academia to review some complex issues. Helping researchers understand the latest developments and theories in his area as soon as possible.

*3.3. querying webpage*
By entering "web + summarize" followed by the URL to receive the main content of that particular page. In that way, what it does is capture the content of the webpage and give the gist of a specific webpage. Similarly, when a user types "web + give me the header + URL," he will have all the headings and subheadings of that particular page. This process extracts formatted text and displays the heading structure as well as their relationships in a sequence. Presently, this is not achievable by the GPT model independently. However, this can be achieved through our system.

*3.4. Comprehensive query*
Our framework has three types of outputs directed at research and information retrieval processes:
a) **GPT-4's Original Response:** With the newest version, it would be one of those machine-type answers it learned from and therefore generated from its training data.
b) **Google Search's Response:** This response incorporates the power of Google Search as it extracts information in real-time, thereby providing data and insights directly from reliable and current sources on the web.

c) **Hybrid Model Response Based on LangChain:** A more fleshed-out reply is the native responses from Google Search, coupled with the major output from GPT-4, synthesized into one coherent whole using the LangChain framework. This synthesized output has been elaborately hand-curated in depicting relevant links next to an exhaustive summary, thus giving enhanced functionality compared with what would come about with just pure use of GPT-4[15].

As mentioned above, the approach is integrated, offering users a rounded understanding of how to embed the benefits offered by AI-created insights, as well as the accuracy and speed of information gathered on the web. The hyperlinks and short abstracts related allow users deeper insights into smooth organization for increased relevance and utility of the research tool.

## 4. Experiments

This section will present two experiments to evaluate the effectiveness of LangChain-based models in domains such as question answering and information retrieval understanding via AI Agents[16]. These results will be compared with the GPT-4o baseline model (temperature = 0). Later, we will provide our main findings in section four.

*3.1. Dataset*

The **Paul Graham Essays Questions-Answers** dataset has 22 questions, and all are derived from essays written by Paul Graham, which the model has to answer. It is intended to assess the efficiency of the developed model in grasping a wide range of text inputs conveying in-depth ideas with textual richness complex in nature.

The **Google Natural Questions** dataset contains questions to which the answers are meant to be derived by extracting actual information out of Wikipedia articles on the fly. It encompasses over 300,000 questions, like "Who was the first Nobel Prize winner in Physics?" and "What is the capital of Australia?", each with a complex answer derived from some corresponding Wikipedia page, a simplified answer in the form of specific names, dates, or locations is also given. This dataset poses a significant test for the model's capacity to comprehend and extract specific information from an extensive corpus, effectively differentiating between pertinent and extraneous content within intricate documents.

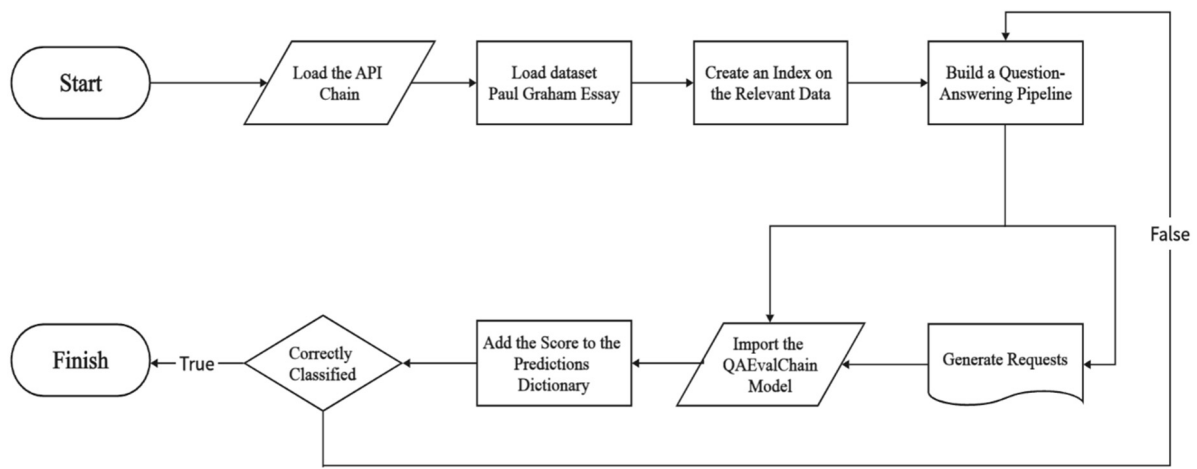

Fig.2 Construction of experiments

*3.2. Implementation*

The performance is tested over two question-answering benchmarks: Paul Graham Essay and Google's Natural Questions. The benchmarks carry out testing of the model in understanding text material with respect to domain-specific reasoning; thus, LangChain-based models are tested for effectiveness in carrying out the tasks within such an area using a collection of texts gathered from the essay set by Paul Graham in pipelines, as presented in **Figure 2**. The Natural Questions benchmark from Google tests the model's ability to do real-world questions over complex information retrieval tasks. This dataset is used as one of the measures of the model's understanding of how to correctly extract answers from similarly large unstructured textual sources, for example, Wikipedia. Following established evaluation protocols, this benchmark shows how well the model can answer a wide range of complex questions under realistic conditions, ensuring it can be relied on to effectively retrieve relevant information.

## 5. Test Results and Discussions

*4.1. Description of specimens*

**Figure 3** presents a real search for YOLOv10, which is a powerful document detection framework that eliminates the requirement for Non-Maximum Suppression during the training phase.

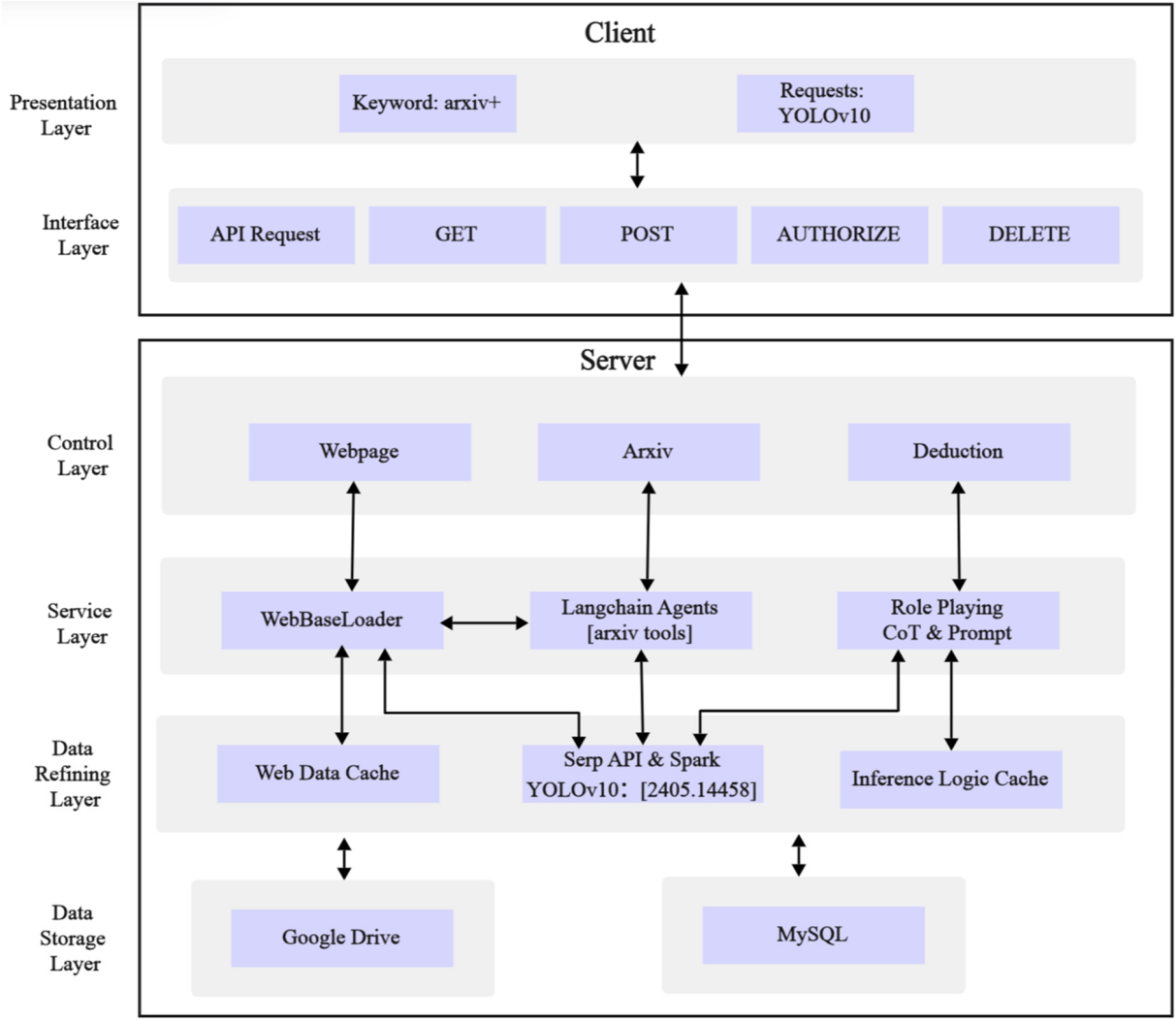

Fig.3 Client-Server Architecture for YOLOv10 Search

The output of the YOLOv10 model should be included in the pipeline so that its performance can be checked in actual comprehensive object detection under a framework of testing, such as Langchain. The benchmark setup will include the development of datasets of different objects under different conditions and then feed that onto the system for performance related to these identification and localization tasks. It would test precision, recall, and mean Average Precision-mAP for performance indicators of model exactitude. It also considered response time and throughput to verify the system's compliance with the requirements of a real-time application.

Due to the implementation of a cache-like mechanism, key information retrieved during each query is recorded and backed up on Google Drive. When subsequent requests are made via the MultiQueryRetriever[17], the relevant content is directly integrated with the previously stored key information and returned to the user. This approach enhances the alignment between the information and the user's needs, while effectively reducing the time delay caused by repeated API calls.

*4.2. Main Results*

a) The question-answering benchmark: Paul Graham Essay

Table 1 Results on Paul Graham Essay Question Answering Dataset

| Model | Accuracy | Precision | mAP |
|---|---|---|---|
| **GPT-3.5 Turbo** | 67.45% | 50.08% | 0.44 |
| **GPT-4o** | 79.09% | 61.23% | 0.52 |
| **BERT** | 85.18% | 63.92% | 0.63 |
| **Prompt-CoT$_{Base}$** | **96.36%** | **89.66%** | **0.87** |

The original GPT-4o, the BERT model, and the one based on LangChain were measured in the Paul Graham Essay Question Answering dataset. It has also been indicated that the incorporation of prompt engineering as well as CoT into the model greatly improved the capability of the model to extract and understand text. In detail, the model based on LangChain achieved a significant performance boost with an accuracy of 96.36%, a precision of 89.66%, and an mAP of 0.87, compared to 79.09% accuracy, 61.23% precision, and 0.52 mAP for GPT-4o.

b) Evaluating on Google's Natural Questions (Long answer Test)

Table 2 Score statistics for the Evaluation Session

| Model | Precision | Recall | F1 Score |
|---|---|---|---|
| **GPT-3.5 Turbo** | 39.11% | 41.82% | 0.35 |
| **GPT-4o** | 49.65% | 47.13% | 0.48 |
| **BERT** | 52.79% | 49.98% | 0.51 |
| **Prompt-CoT$_{Base}$** | **67.12%** | **60.03%** | **0.63** |

In the second experiment, we evaluated our improved model, based on LangChain, on Google's Natural Questions dataset. As shown in Table 2, our enhanced model demonstrates a substantial improvement in performance over models like GPT-4o and BERT, meanwhile, it is rather more comparable to human performance on the current task, particularly in terms of recall and overall answer accuracy. In detail, the F1 score of Prompt-CoT$_{Base}$ reaches 0.63, which is significantly higher than GPT-4o's 0.48 and BERT's 0.51, indicating our model's enhanced ability to provide accurate and relevant long answers.

## 6. Conclusion

In this paper, we combine different models into the LangChain framework, greatly expanding its functionality compared to the original GPT-4o, which was limited in its capabilities. This integration enables the system to query arXiv papers, web pages, and handle more sophisticated queries. Through prompt engineering with chain-of-thought reasoning, we fused Google Search with GPT-4o to yield a much more accurate and timely question-answering model. The experiments and evaluations indicate that such an integration indeed improves response accuracy, with a much deeper contextual understanding and reasoning of the model. These findings highlight the potential for further refinement, particularly through the inclusion of more diverse data sources and the continued optimization of CoT mechanisms, especially for multimodal reasoning questions, to advance the performance of AI-driven information retrieval systems in future research.

**Acknowledgments**
Wenqi Guan and Yang Fang contributed equally to this work and should be considered co-first authors.